\documentclass[10pt]{wlscirep}
\usepackage{wrapfig}
\usepackage{subcaption}
\usepackage{amsmath}
\usepackage{color}
\usepackage{nicefrac}
\usepackage[capitalize]{cleveref}
\usepackage{bm}

\newcommand{\eg}{\emph{e.g.,~}}
\newcommand{\ie}{\emph{i.e.,~}}

\newcommand{\ket}[1]{\vert#1\rangle}

\newcommand{\expect}[1]{\langle #1\rangle}
\newcommand{\dg}{^\dagger}

\newcommand{\dd}{{\rm d}}

\title{Data-driven learning of non-Markovian quantum dynamics}

\author{Samuel Goodwin$^{1,3}$, Brian K. McFarland$^{2}$, Manuel H. Mu\~noz-Arias$^{3}$, Edward C. Tortorici$^{2}$, Melissa C. Revelle$^{2}$, Christopher G. Yale$^{2}$, Daniel S. Lobser$^{2}$, Susan M. Clark$^{2}$, Mohan Sarovar$^{3}$\\
	\textit{$^{1}$~Department of Physics and Astronomy, Center for Quantum Information and Control, University of New Mexico,
		Albuquerque, New Mexico 87106, USA}
	\\
	\textit{$^{2}$~Sandia National Laboratories, Albuquerque, NM, 87185, USA}
	\\
	\textit{$^{3}$~Quantum Algorithms and Applications Collaboratory, Sandia National Laboratories, Livermore, CA 94550, USA}
	\\
	\textit{sgoodwin2@unm.edu}
}

\begin{abstract}
		Fault-tolerant quantum computing requires extremely precise knowledge and control of qubit dynamics during the application of a gate. We develop a data-driven learning protocol for characterizing quantum gates that builds off previous work on learning the Nakajima-Mori-Zwanzig (NMZ) formulation of open system dynamics from time series data, which allows detailed reconstruction of quantum evolution, including non-Markovian dynamics. 
        We demonstrate this learning technique on three different systems: a simulation of a qubit whose dynamics are purely Markovian, a simulation of a driven qubit coupled to stochastic noise produced by an Ornstein-Uhlenbeck process, and trapped-ion experimental data of a driven qubit whose noise environment is not characterized ahead of time. Our technique is able to learn the generators of time evolution, or the NMZ operators, in all three cases and can learn the timescale in which the qubit dynamics can no longer be accurately described by a purely Markovian model. Our technique complements existing quantum gate characterization methods such as gate set tomography by explicitly capturing non-Markovianity in the gate generator, thus allowing for more thorough diagnosis of noise sources. 
	\end{abstract}
    
\begin{document}
	
	\maketitle

\section*{Introduction}
		The biggest obstacle to successfully building a fault-tolerant quantum computer is noise.  Current quantum hardware is classified as Noisy Intermediate-Scale Quantum (NISQ) because qubits in this hardware are too susceptible to noise from the environment and other qubits \cite{nisq}.  From this perspective, all qubits are open quantum systems that have interactions with an environment that must be accounted for to accurately understand their behavior. Precisely characterizing these open quantum systems, understanding sources of noise, and ultimately, overcoming them is critical to transitioning from currently available quantum devices to fault-tolerant quantum computers. 

        Most existing methods for characterization of qubits and quantum gates capture the overall map a system undergoes during a time period, usually in the form of a \emph{quantum channel} or \emph{process matrix} \cite{Nielsen2021-ex}. While such characterization is useful, it often does not allow for detailed understanding of the dynamics of a quantum system during the time period of interest, \eg during the application of a gate.  Having a precise understanding of these intermediate dynamics is often essential to identifying and mitigating the sources of noise affecting the quantum system. In cases where the generator of evolution is estimated with existing techniques, it is assumed to be Markovian \cite{Nielsen2021-ex} or a specific parameterized form of non-Markovian evolution \cite{Vinas2025}, and therefore only accurate up to these model assumptions. An alternative approach to direct characterization of qubits is to model their dynamics through numerical simulations. While such modeling aims to capture dynamics in detail, this approach is susceptible to model accuracy and tractability issues. Hamiltonian or phenomenological noise models like the Lindblad master equation or the stochastic noise channel are efficient in time and memory, but rely on uncontrolled approximations or assumptions about the environment. Detailed first-principles modeling properly captures all of the physics and can often fit to experimental data, but this type of modeling is very resource-intensive \cite{first-principles} and often intractable for realistic qubit systems.  
                
        In this work we present a model-free, data-driven approach to understanding the dynamics of a quantum system to unprecedented detail, including the precise characterization of \emph{non-Markovian dynamics}. Non-Markovian dynamics are the result of non-negligible memory effects in the environment coupling to a quantum system of interest, and are seen in most qubit platforms. Understanding the timescales of this non-Markovianity is extremely important to the design of effective control and error correction protocols \cite{White2020, control-and-correction}. Being model-free and data-driven, our approach does not rely on approximations of the environment or intractable first-principles models, and thus fills a critical gap in the characterization and modeling of quantum systems.

        Our approach is based on the Nakajima-Mori-Zwanzig (NMZ) formulation of open quantum dynamics \cite{nmz, bp}, which formulates the dynamics of a quantum system coupled to an arbitrary environment as an integro-differential equation:
        \begin{align}
        \frac{d}{dt}\mathbf{g}(t) = \mathbf{M}\cdot\mathbf{g}(t) - \int_0^t \mathbf{K}(s)\cdot\mathbf{g}(t-s)ds + \mathbf{F}(t)
        \label{eq:NMZ_general}
        \end{align}
        where $\mathbf{g}(t)$ is a vector of observable expectations for the quantum system of interest, $\mathbf{M}$ is the \emph{Markov transition matrix}, $\mathbf{K}(t)$ is the \emph{memory kernel}, and $\mathbf{F}(t)$ is a time-dependent \emph{Langevin noise} term. Note that we set $\hbar=1$ throughout.  Being an \emph{exact} representation of the dynamics, this formulation is in general, no more tractable than any other formulation of general open quantum system dynamics -- it trades off the complexity of modeling the environment degrees of freedom for the complexity of modeling the memory kernel and Langevin noise terms. However, for quantum systems engineered to be good qubits we expect the environmental degrees of freedom to have finite and short correlations, and thus for the memory kernel to be fast decaying. In addition, we argue in the Methods section that the Langevin noise term should be zero for all time in realistic settings in quantum computing. Accounting for these factors, the NMZ equation can be a significantly more parsimonious description of open quantum dynamics than other treatments, an observation that we will validate in the Results section below.

        While the NMZ equation has been known for decades, it has not found widespread use in modeling quantum computing systems. Part of the reason for this is that deriving the Markov transition matrix and memory kernel for a system is difficult, even if the system-environment interaction and environment dynamics are known. To overcome this difficulty in this work, we use a data-driven approach and \emph{learn} these operators from time series data. We utilize recent results from Lin \emph{et al.}, where this task is formulated as a linear regression problem that admits an extremely efficient solution \cite{lin2021, lin2023}. We demonstrate the application of this learning algorithm in our setting and show how the learned operators can be used to formulate a predictive model of qubit dynamics and also interpreted to gain information about the qubit dynamics. 

        We conclude the Introduction by discussing other work relevant to our results. 
        Perhaps the most mature data-driven method for estimating quantum dynamics is gate set tomography \cite{Nielsen2021-ex}, which has been extensively deployed to characterize experimental hardware. However, it is limited to the characterization of quantum computing gates and further, assumes that gates are generated by Markovian generators, something we explicitly avoid. Several works have demonstrated estimation of quantum dynamics using neural networks, including characterization using black box models \cite{Mazza,Dressel,rnn} and Markovian parameterizations of dynamical generators \cite{Cemin,Cemin_NISQ,Annupam}. Typically, these neural network methods require large amounts of training data in order to produce accurate predictive models, especially in the case of black box models. Parametrized generators of Markovian quantum dynamics can also be estimated from time series data using algebraic inversion \cite{Zhang_2014,Zhang_2015}. 
        An early attempt at characterizing \emph{non-Markovian} dynamics developed a \emph{tensor transfer} method to reconstruct dynamical generators, which can be related to the NMZ equation \cite{OQT}. While the transfer tensors do resemble our parametrization of the Markov transition matrix and memory kernel, they are iteratively constructed from the dynamical maps, which must be either known or learned in advance, for instance via process tomography, limiting the technique to systems where this initial step can be performed efficiently. See Li \emph{et al.} \cite{li2025probingnonmarkovianqubitnoise} for a similar approach that requires full characterization of the dynamical process, \eg via process tomography, before fitting to a ``post-Markovian'' master equation that capture non-Markovian effects. An alternative direction for characterizing quantum dynamics proceeding via deep learning is the STEADY method \cite{steady}, which can fit to non-Markovian models such as the NMZ equation. 
        Despite being physics-informed, the STEADY method, being deep learning-based, is likely to have heavier training data requirements than the technique we present here, which relies on efficient linear regression. Finally, recent work by Reddy aims to learn an alternative representation of open quantum dynamics, the time-convolutionless master equation \cite{Reddy-2025}. This work also assumes certain parameterized forms for the evolution in order to facilitate efficient learning, which reduces the generality of the learned models. We note that Reddy mentions the NMZ equation as a possible model for open quantum dynamics, but sets it aside due to concerns about intractability of learning. Our work addresses these concerns and demonstrates that learning the NMZ operators can be very tractable in realistic situations, including for learning from noisy experimental data.

\section*{Results}
In the following, we demonstrate the application of our data-driven characterization approach to learn the dynamics of a qubit under the application of a noisy quantum gate, using simulated data and experimental data. We focus on the evolution of a single qubit in this work to clearly demonstrate the techniques and interpret the results. However, it should be noted that our approach generalizes trivially to larger systems.  The general learning algorithm based on time series of the observable vector $\mathbf{g}(t)$ is detailed in the Methods section. 

The time-dependent state of a single qubit $\rho(t)$ can be written as $\rho(t) = \frac{1}{2}(\mathbf{I} + \mathbf{r}(t)\cdot\mathbf{\sigma})$, where $\mathbf{r}(t)$ is the \emph{Bloch vector} of the qubit and $\mathbf{\sigma}$ is the Pauli vector $\mathbf{\sigma} = (\sigma_x, \sigma_y, \sigma_z)$.  Our observables of interest are the elements of the Bloch vector, $r_{\alpha}(t) = \text{Tr}(\rho(t)\sigma_{\alpha}) \equiv \expect{\sigma_{\alpha}}(t)$, which completely characterize the qubit.  We also include a bias term in order to capture non-unital dynamics, which makes our observables vector $\mathbf{g}(t) = (1,\expect{\sigma_x}(t),\expect{\sigma_y}(t),\expect{\sigma_z}(t))^{\sf T}$.

The general learning algorithm learns the operators $\mathbf{M}, \mathbf{K}, \mathbf{F}$ for a discretized version of the NMZ equation since the data fed into the learning algorithm is taken at discrete times.

\begin{align}
	\mathbf{g}((k+1)\Delta) = \sum_{l=0}^k \bm{\Omega}_{\Delta}^{(l)} \cdot \mathbf{g}((k-l)\Delta) + \mathbf{W}_k
\end{align}

\noindent In the discretized form of the NMZ equation, $\Delta$ is the amount of time between sampled points in the trajectory, $k \in \mathbb{Z}$ is a timestep index such that $t = k\Delta$, $\bm{\Omega}_{\Delta}^{(l)}$ is the $l$-th order NMZ operator that encodes $\mathbf{M}$ or $\mathbf{K}$ in the discretized form of the NMZ equation, and $\mathbf{W}_k$ are the Langevin terms that discretize the noise $\mathbf{F}(t)$.  We prove in the Methods section the Langevin noise terms do not make a significant contribution to the system's dynamics, so we can assume the noise terms are zero.  There is no straight-forward way to convert between $(\mathbf{M},\mathbf{K})$ and $\bm{\Omega}_{\Delta}^{(l)}$ in general, but if $\Delta$ is very small, then we can approximate \cite{lin2021} $\bm{\Omega}_{\Delta}^{(0)} \approx \mathbf{I} + \mathbf{M}\Delta$ and $\bm{\Omega}_{\Delta}^{(l)} \approx \Delta^2\mathbf{K}(l\Delta)$ for $l > 1$.

\subsection*{Simulation: Markovian evolution}
First we benchmark our technique on the simplest open quantum system dynamics -- Markovian evolution. A fairly general generator of autonomous, Markovian evolution for a single qubit is given by the master equation:
\begin{align}
    \frac{\dd \rho(t)}{\dd t} = -i[\omega_x \sigma_x + \omega_y \sigma_y + \omega_z \sigma_z, \rho(t)] + \Gamma_x \mathcal{D}[\sigma_x]\rho(t) + \Gamma_y \mathcal{D}[\sigma_y]\rho(t) + \Gamma_z \mathcal{D}[\sigma_z]\rho(t) + \gamma_+ \mathcal{D}[\sigma_+]\rho(t) + \gamma_- \mathcal{D}[\sigma_-]\rho(t),
    \label{eq:ME_general}
\end{align}
where $\rho(t)$ is the qubit density matrix, $\sigma_{\pm} = \nicefrac{1}{2}(\sigma_x \pm i \sigma_y)$, and $\mathcal{D}[A]\rho \equiv A\rho A\dg -\nicefrac{1}{2}A\dg A \rho - \nicefrac{1}{2}\rho A\dg A$. All other parameters in this equation are rates of various coherent and incoherent processes. While more general generators of Markovian evolution are possible, the above captures all the components of physically realistic Markovian evolution. In terms of the terminology established in Ref. \cite{PRXQuantum.3.020335}, certain rarely-seen stochastic Pauli correlation generators and active generators are omitted from the above description.  

By integrating Eq. \eqref{eq:ME_general} for a small timestep $\Delta$, we can derive the general finite-time Markov transition matrix for this system:

\begin{align}
	\bm{\Omega}_{\Delta}^{(0)} =
				\left(\begin{smallmatrix}
					1 & 0 & 0 & 0 \\
					0 & 1 - 2\Delta(\Gamma_y + \Gamma_z) - 0.5\Delta(\gamma_+ + \gamma_-) & -2\Delta\omega_z & 2\Delta\omega_y  \\
					0 & 2\Delta\omega_z & 1 - 2\Delta(\Gamma_x + \Gamma_z) - 0.5\Delta(\gamma_+ + \gamma_-) & -2\Delta\omega_x  \\
					\Delta(\gamma_+ - \gamma_-) & -2\Delta\omega_y & 2\Delta\omega_x & 1 - 2\Delta(\Gamma_x + \Gamma_y) - \Delta(\gamma_+ + \gamma_-)
				\end{smallmatrix}\right)
                \label{eqn:markov_generator}
\end{align}

Note that this transition matrix operates on a Bloch vector augmented with a constant term to capture the non-unital dynamics of the general master equation \cref{eq:ME_general}, \ie $\mathbf{g}(t) = (1, \expect{\sigma_x}(t), \expect{\sigma_y}(t), \expect{\sigma_z}(t))^{\sf T}$. We simulate time series data under this dynamics by repeatedly applying this transition matrix to 10 random (pure) initial states. The parameters chosen are: $\omega_z=1$, $\omega_x=\omega_y=0$, $\Gamma_x=0.1$, $\Gamma_y=\Gamma_z=0$, $\gamma_-=0.4$, and $\gamma_+=0$. The time series are generated for a total time of $T=20$ (arb. units) with timesteps $\delta=10^{-3}$. This data is then downsampled for training NMZ models at a rate $\Delta = 100\delta$.  The NMZ model is trained using a \emph{leave one out cross validation} (LOOCV) approach, where the model is trained on 9 of the 10 time series and then tested on the 10th series. This is repeated 10 times, with the test time series being different each time, resulting in a total of 10 trained NMZ models. We do not assume Markovianity, and so we train the NMZ models to learn a small memory kernel, which we expect to worsen the performance of the models since a non-zero memory kernel is just adding noise. 

We assess the quality of the learned model in two ways: by comparing the learned Markov transition matrix to the true version, and by computing the root mean squared error (RMSE) in the predicted time evolution (averaged across the 10 cross validations). The RMSE for a single model is computed by squaring the deviation between the predicted and true values for each observable and across all time steps, and then taking the square root:
\begin{align}
	RMSE = \sqrt{\frac{1}{K} \sum_{j=\{x,y,z\}} \sum_{k=0}^{K-1} [\expect{\sigma_j(k\Delta)} - \expect{\hat{\sigma}_j(k\Delta)}]^2},
	\label{eq:rmse}
\end{align}
where the predicted values are denoted by the hat, and $K=T/\Delta$. In the LOOCV setting, we will often average the RMSE across all learned models, and this averaged quantity is denoted $\overline{\text{RMSE}}$.

\begin{figure*}
        \centering
			\begin{subfigure}{\textwidth}
				\includegraphics[width=0.9\textwidth]{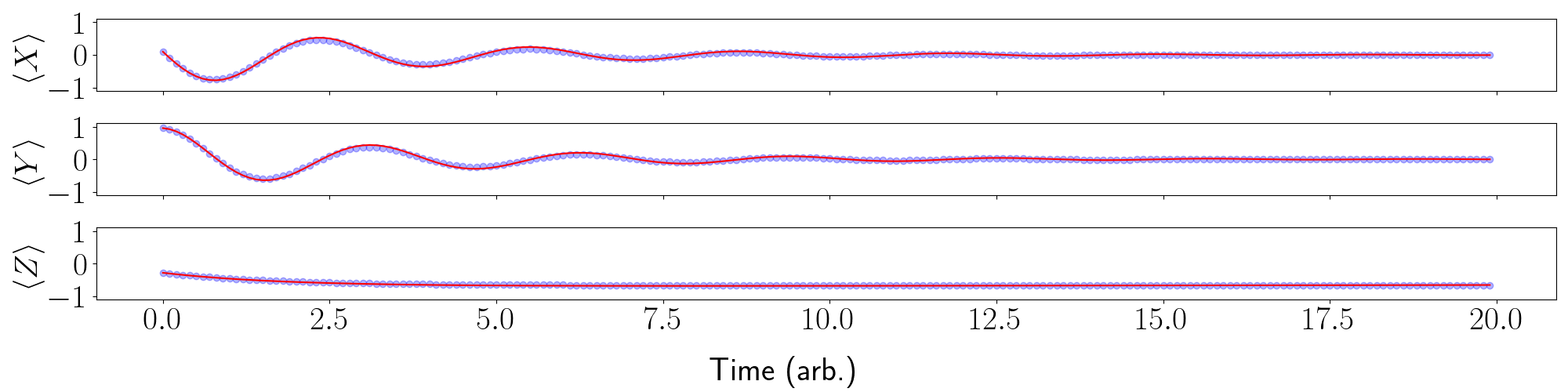}
				\vspace{5pt}
				\caption{}
			\end{subfigure}
			\begin{subfigure}{0.4\textwidth}
				\includegraphics[width=\textwidth]{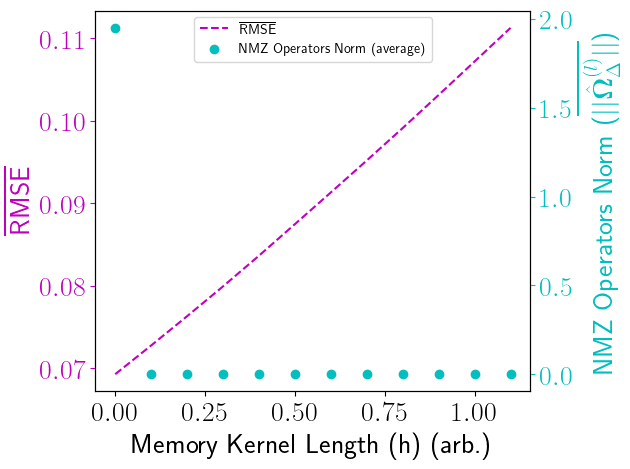}
				\caption{}
			\end{subfigure}
			\begin{subfigure}{0.4\textwidth}
				\includegraphics[width=\textwidth]{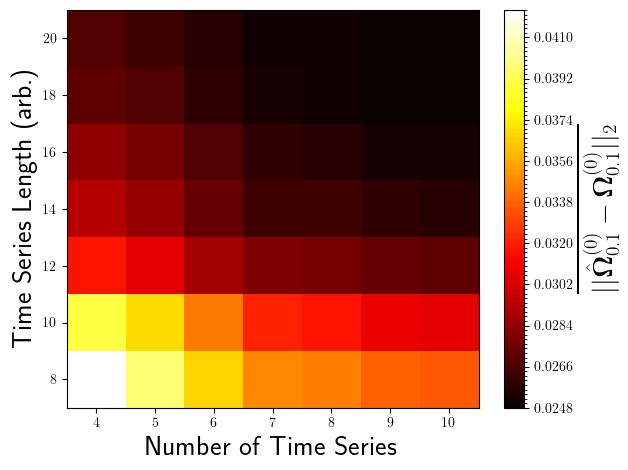}
				\caption{}
			\end{subfigure}
			
			\caption{Demonstration of the NMZ learning algorithm on a qubit undergoing Markovian dynamics (under \cref{eq:ME_general} with parameters defined in the text). \textbf{(a)} shows how an NMZ model is able to accurately predict the dynamics of the simulated qubit. The blue points are the simulated qubit Bloch vector data, and the red lines are predictions from the trained NMZ model. \textbf{(b)} shows predictive performance ($\overline{\text{RMSE}}$, dashed magenta line) and the averaged norms of the learned NMZ operators ($\overline{\bm{\hat{\Omega}}_\Delta^{(l)}}$, cyan markers) as a function of memory kernel length. \textbf{(c)} shows the average error in the predicted Markov matrix, $\overline{||\hat{\bm{\Omega}}_{0.1}^{(0)} - \bm{\Omega}_{0.1}^{(0)}||_2}$, as a function of the length and number of time series used for training.  \label{fig:ME}}
		\end{figure*}

\cref{fig:ME}(a), which shows the actual and predicted Bloch vector for a test initial state, illustrates the predictive power of the learned model. The discrepancy between the true data and the predictions are hardly resolvable. To more systematically capture the prediction error, \cref{fig:ME}(b) shows the RMSE to true observable evolution (averaged across cross validation samples) of the learned model, and the average $L2$ norm of the learned NMZ operators, as a function of memory kernel length, $h = l\Delta$. As expected, increasing the memory kernel length only increases the predictive error. Moreover, the norms of the predicted NMZ operators, $\hat{\bm{\Omega}}_{\Delta}^{(l)}$, for $l>0$ are extremely small.  Both of these facts are strong evidence that the true dynamics is Markovian. 

Finally, we examine the value of the Markov matrix learned by our procedure. This learned matrix is slightly different in each cross validation, but an example learned by one model is:
\begin{align*}
                \small
				\hat{\bm{\Omega}}_{0.1}^{(0)} =
				\begin{pmatrix}
					1.01 & -5.0\times10^{-4} & 2.4\times10^{-4} & 9.4\times10^{-3} \\
					6.44\times10^{-6} & 0.966 & -0.194 & 6.03\times10^{-6} \\
					1.70\times10^{-6} & 0.194 & 0.946 & 7.97\times10^{-6} \\
					-0.04 & 3.3\times10^{-4} & -1.6\times10^{-4} & 0.940
				\end{pmatrix}
\end{align*}
The error in this learned matrix averaged across all models is $\overline{||\hat{\bm{\Omega}}_{0.1}^{(0)} - \bm{\Omega}_{0.1}^{(0)}||_2} = 0.025$. This error is dictated by the number of time series used for training and the total time of each time series in the training data set. This claim is validated by Fig. \cref{fig:ME}(c), which shows the average error in the learned Markov matrix as a function of the length and number of time series used for training. It is important to train the model on time evolutions with a variety of initial states and for long  enough evolution times that transients subside. 

Given the model of Markovian evolution given in \cref{eqn:markov_generator}, we can interpret the learned Markov matrices and extract physical parameters for the dynamics. 
				
		\begin{figure*}[!t]
			\begin{subfigure}{0.3\textwidth}
				\includegraphics[width=\textwidth]{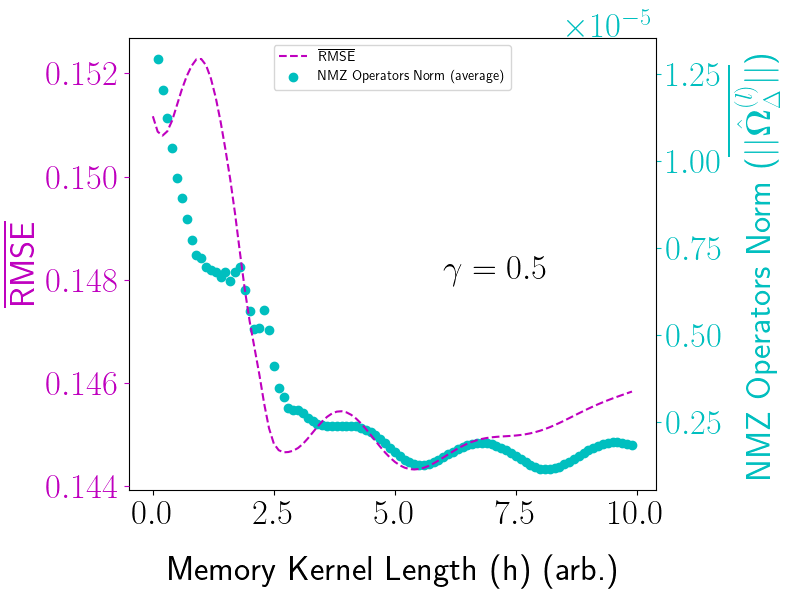}
				\includegraphics[width=\textwidth]{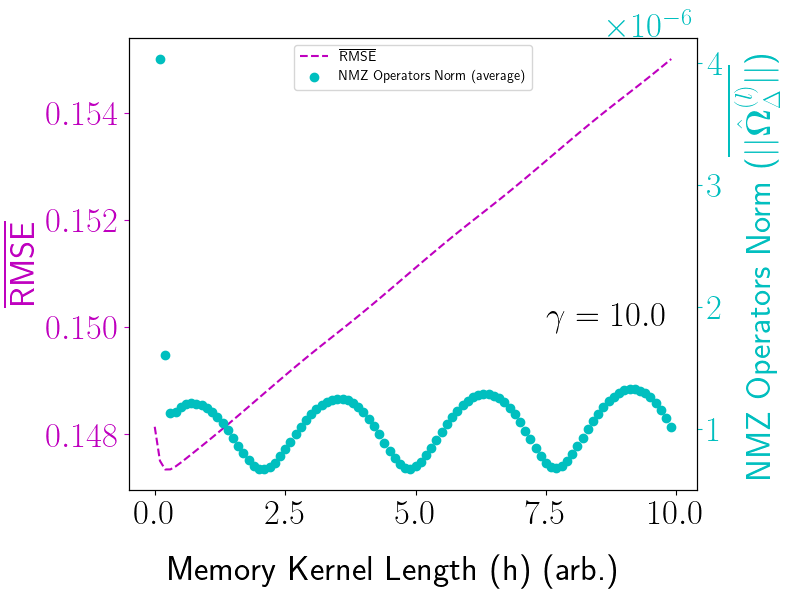}
				\label{fig:ou_error_plot}
				\caption{}
			\end{subfigure}
			\begin{subfigure}{0.65\textwidth}
				\includegraphics[width=\textwidth]{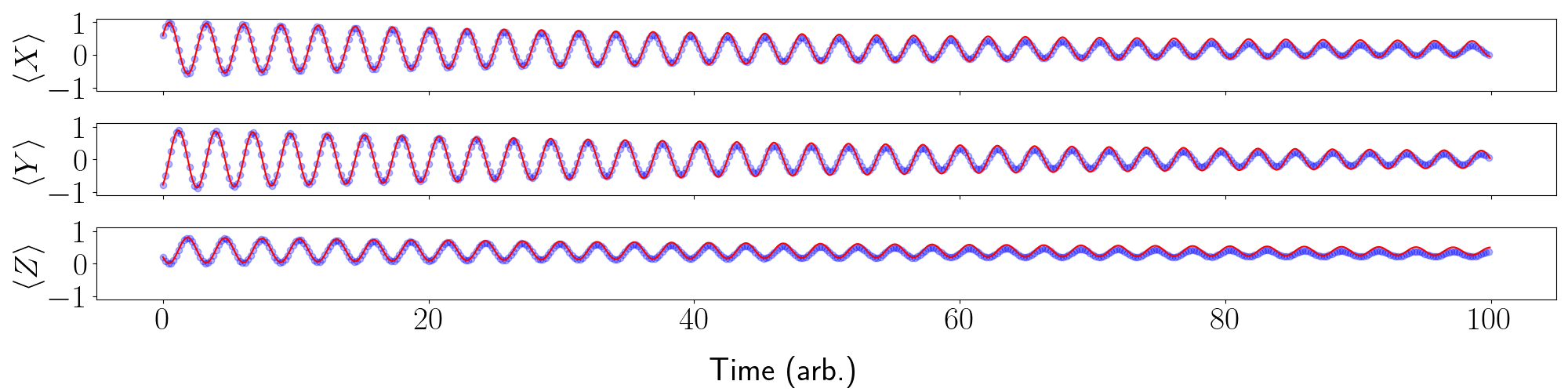}
				\vspace{20pt}
				\includegraphics[width=\textwidth]{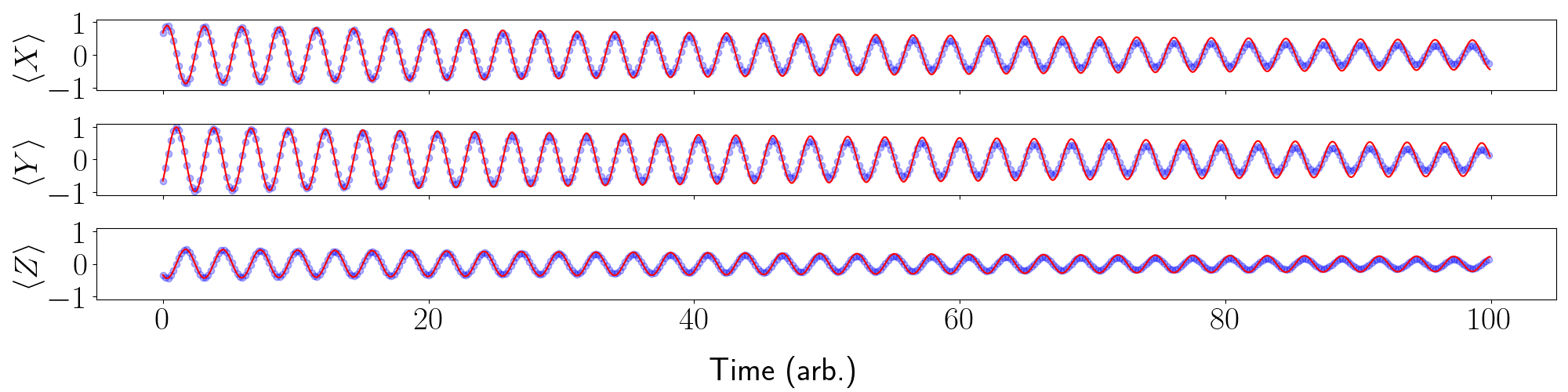}
				\label{fig:ou_comparisons}
				\vspace{-10pt}
				\caption{}
			\end{subfigure}
			
			\caption{NMZ operator learning for simulated driven qubit with OU noise. \textbf{(a)} The $\overline{\text{RMSE}}$ of the learned NMZ model predictions (dashed magenta line) and the averaged L2 norms of the learned NMZ operators (cyan  markers) as a function of memory kernel length for $\gamma = 0.5$ (strongly non-Markovian) and $\gamma = 10.0$ (weakly non-Markovian). \textbf{(b)} shows an example of predictions from the learned model when applied to an unseen initial state. The top set of plots shows predictions for the $\gamma = 0.5$ data when the memory kernel length is truncated at $h^\star = 5.0$ $(arb.)$. The bottom set of plots shows predictions for the $\gamma = 10.0$ data where the memory kernel length is truncated at $h^\star = 0.2$ $(arb.)$. 
            The blue points are the simulated data, and the red lines corresponds to the predictions made by the learned NMZ models. }
            	\label{fig:ou}
		\end{figure*}
\subsection*{Simulation: non-Markovian evolution}
Next, we demonstrate the NMZ learning algorithm on simulated data of non-Markovian evolution. We simulate the dynamics of a single qubit undergoing a coherent rotation and subject to stochastic noise along an orthogonal axis. The Hamiltonian for the qubit evolution is $H(t) = \omega_z \sigma_z + \eta(t) \sigma_x$, where $\eta(t)$ is an Ornstein-Uhlenbeck (OU) stochastic process, following the evolution equation $\dd \eta(t) = -\gamma(\eta(t)- \mu)\dd t + \sigma \dd W_t$, with $\dd W_t$ being a Wiener increment. The OU process is a mean-reverting noise process, with $\mu$ being the mean, $\gamma>0$ the rate of reversion to the mean, and $\sigma>0$ the rate of fluctuations. Values of the OU process at different times $t, t'$ are correlated as $\left(\nicefrac{\sigma^2}{2\gamma}\right) e^{-\gamma|t - t'|}$. 
This correlation implies the evolution of the qubit driven by this noise is non-Markovian, with the Markovian limit given by $\gamma \rightarrow \infty$. For our simulations, we choose the parameters $(\gamma=0.5, \sigma=0.1)$ (strongly non-Markovian) and $(\gamma=10.0, \sigma=1.0)$ (weakly non-Markovian). In both simulations, we set $\mu = 0.5$.  In order to simulate average qubit dynamics under this stochastic noise, we used the diffusive hierarchical equations of motion (DHEOM) developed in Ref. \cite{mohan}. As with the Markovian example, we simulate dynamics starting from 10 random (pure) initial states. The $\gamma = 0.5$ data is simulated with timestep $\delta = 10^{-3}$ for a total time of $T = 200$ (arb.). This data is then downsampled for training NMZ models at sampling timestep $\Delta = 100\delta$. The $\gamma = 10.0$ data was simulated with timestep $\delta = 10^{-2}$ for a total time of $T = 200$ (arb.). This data is then downsampled for training NMZ models at sampling timestep $\Delta = 10\delta$. Again, we use a LOOCV approach to training: train the NMZ model on 9 randomly sampled time series and test its predictive power using the remaining time series, and then repeat. 

Figure 2 shows the performance of the learned models. Fig. 2(a) shows the average RMSE in prediction and average value of the $L2$ norms of NMZ operators as a function of memory kernel length, $h=\Delta l$, for the $\gamma=0.5$ and $\gamma=10.0$ data. In contrast to the Markovian example in the previous section, we see that the error on predicting the $\gamma=0.5$ data decreases rapidly as memory kernel length increases and eventually reaches a constant error. This rapid decreases in prediction error with increasing memory kernel length also shows this system is strongly non-Markovian when $\gamma=0.5$.  We also see the prediction error for the $\gamma=10.0$ data mostly increases as memory kernel length increases, which indicates the system is weakly non-Markovian when $\gamma=10.0$. These trends in average RMSE in prediction are reflected in the behavior of the average norm of NMZ operators also, $||\hat{\bm{\Omega}}_\Delta^{(l)}||$. In the $\gamma=0.5$ case, the norms decay slower with $l$ than for $\gamma=10.0$.  We removed the norms $||\hat{\bm{\Omega}}_\Delta^{(0)}||$ from these plots because they are orders of magnitude larger than the norms $||\hat{\bm{\Omega}}_\Delta^{(l \geq 1)}||$.

The decay of the average RMSE and the average norm of the learned NMZ operators indicate that the memory kernel can be truncated at $h^\star\sim 5$ (arb.) for $\gamma=0.5$, and 
at $h^\star\sim 0.2$ (arb.) for $\gamma=10.0$ without sacrificing predictive performance.
These truncations indicate the degree of non-Markovianity in the qubits dynamics. Fig. \ref{fig:ou}(b) shows how the learned NMZ models with these truncated memory kernels are able to predict the observables of the qubit. 

In the Supplementary Material, we also show how the matrix elements of the learned NMZ matrices ($\hat{\bm{\Omega}}_\Delta^{(l)}$) behave as a function of the memory kernel length. All elements in these matrices decay to zero, which is another criterion for deciding when to truncate the memory kernel length in our learned NMZ models without sacrificing predictive accuracy. 
				
\begin{figure*}[t!]
    \begin{subfigure}{0.3\textwidth}
		\includegraphics[width=\textwidth]{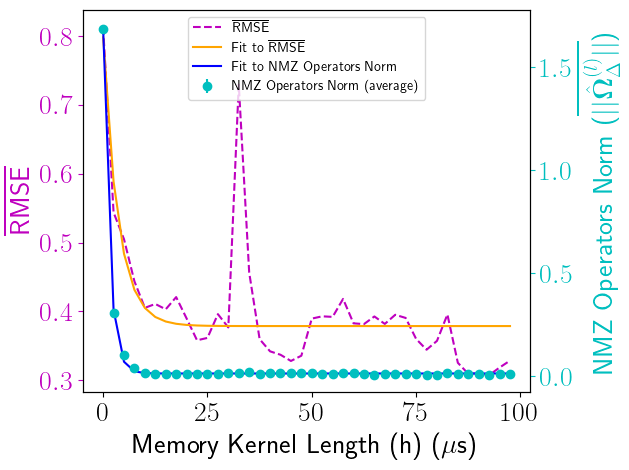}
		\label{fig:qscout_error_plot}
		\caption{}
	\end{subfigure}
	\begin{subfigure}{0.7\textwidth}
		\includegraphics[width=\textwidth]{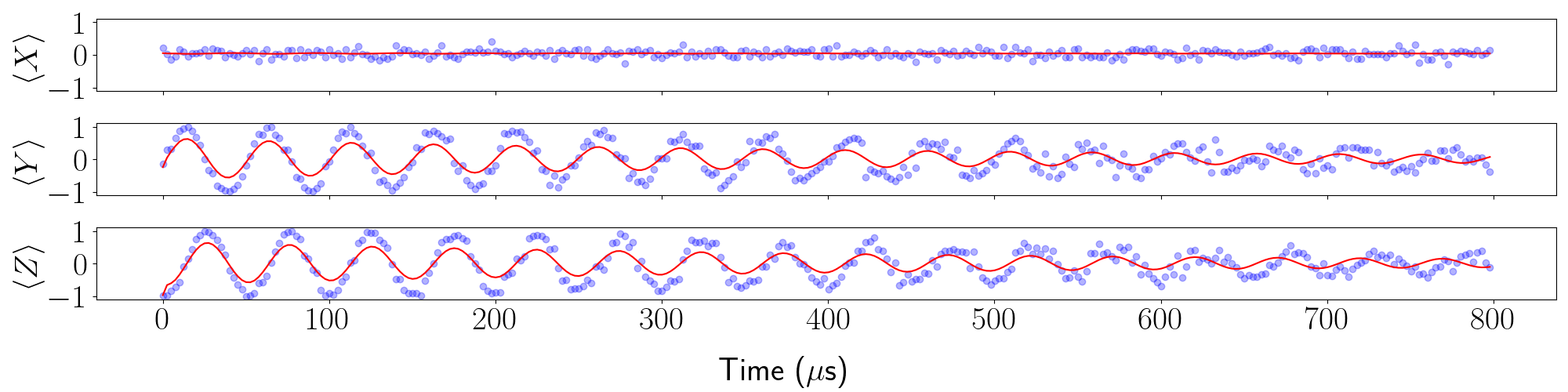}
		\label{fig:qscout_comparisons}
		\vspace{4pt}
		\caption{}
	\end{subfigure}
	\caption{A continuous Rabi drive is applied to a single qubit causing it to rotate about its x-axis for 5000 $\mu$s. \textbf{(a)} $\overline{\text{RMSE}}$ of the learned NMZ model predictions (dashed magenta line) and averaged L2 norms of the learned NMZ operators (cyan markers) as a function of memory kernel length. The orange line is a fit of $\overline{\text{RMSE}}$ to an exponential function, which is $\overline{\text{RMSE}} = 0.35e^{-8.73 (2.5 \mu s)l}+0.34$, where $l$ is the memory kernel length index. The blue line is a fit of the average NMZ operator norms to an exponential function, which is $\overline{|| \hat{\bm{\Omega}}_{2.5 \mu s}^{(l)} ||} = 1.67e^{-0.68 (2.5 \mu s) l}+0.01$.  Guided by the decay of the operator norms, we truncate the memory kernel at $h^\star = 22.5 \mu$s. \textbf{(b)} Predictions from one of the learned NMZ models for an unseen initial state. The blue datapoints are the measured Pauli expectations, and the red line corresponds to the predictions made by the learned NMZ model.
	\label{fig:qscout}}
\end{figure*}

\subsection*{Experiment: learning trapped-ion dynamics}
Experiments were performed on the Quantum Scientific Computing Open User Testbed (QSCOUT)\cite{qscout_2021}, a trapped ion quantum processor located at Sandia National Laboratories. In the experiments, a single ion is trapped, Doppler cooled, and prepared in $\ket{0}$. Tailored Raman pulses then perform rotations on the qubit state to achieve the desired initial state for the measurements. Following this state preparation step, we drive Rabi oscillations on the carrier transition about the $\sigma_x$ axis. The duration of the Raman drive pulses are swept. Sequential measurements of the qubit projections along the $\sigma_x$, $\sigma_y$, and $\sigma_z$ directions are made for each drive pulse duration where $\pi/2$ rotations $Ry(\pi/2)$ and $Rx(\pi/2)$ are applied before detection to measure $\sigma_x$ and $\sigma_y$ projections. This state preparation and measurement procedure is repeated $M$ times (per $\sigma_\alpha$ projection and drive duration) to estimate the Bloch vector expectations, $\expect{\sigma_x}, \expect{\sigma_y}$ and $\expect{\sigma_z}$.  In the long-time driven dynamics experiment, $M = 50$, and in the Pauli-X gate experiment, $M = 400$.  The initial state preparation pulses vary in duration between $12$ and $25$ $\mu$s, and the pulses prior to measurement have duration $12.5$ $\mu$s.

\subsubsection*{Long-time driven dynamics}
First, we characterize the trapped-ion dynamics for long-time drives. After cooling, we prepare the trapped ion qubit in a random initial state by applying a random single qubit $U3$ rotation, and then apply a Rabi drive for up to 5000 $\mu$s. To capture the time series data the Bloch vector of the qubit was measured every 2.5 $\mu$s. This procedure was repeated for a total of 10 random initial states, and these 10 time series were used to train NMZ model using LOOCV, as with the simulation data in previous sections. 

 We test model performance by giving the NMZ model the (ideal) initial state of the test time evolution and then have it predict the Bloch vector of the qubit until $T = 800$ $\mu$s because this is approximately when a steady state is reached.  We evaluate model performance by calculating the RMSE for predicting the test time series, averaged across the 10 trained models.  The RMSE was calculated between the predicted and measured Bloch vector time evolution data within this timeframe.  In order to observe how memory kernel length affects model performance, we repeated these predictions with various memory kernel lengths from a memory kernel of 0 $\mu$s (Markovian model) to 100 $\mu$s in increments of 2.5 $\mu$s.  The plot shown in \cref{fig:qscout}(a) is the average RMSE and NMZ operators norms across all 10 learned NMZ models.  We see the predictive error and the NMZ operator norms quickly decay after introducing a small memory kernel length, indicating this trapped-ion dynamics is non-Markovian when driven at these long timescales. Both the average RMSE and average NMZ operator norm as a function of memory kernel length can be fit to exponential decays, as described in the caption. Guided by the decay of the NMZ operator norms we truncate the memory kernel length at $h^\star=22.5$ $\mu$s.  \cref{fig:qscout}(b) shows the predictive performance of a learned NMZ model with $h^\star=22.5$ $\mu$s memory kernel on the test time series. 

The peak in $\overline{\text{RMSE}}$ in \cref{fig:qscout}(a) around $h=30$ $\mu$s is evidence of instability of the learned NMZ operators due to shot noise in the time traces. This noise can lead to numerical errors that result in unphysical predicted dynamics (\eg Pauli expectations greater than one). Such instabilities in the simple linear regression-based learning procedure used here motivate more advanced learning techniques, and we return to this in the Discussion section. 

We now comment on the source of non-Markovian dynamics identified by our NMZ models. 
Common sources of noise in trapped ions are fluctuations in the driving lasers and motional heating. We rule out the former as the source of the observed non-Markovianity by measuring the spectral density of the amplitude and phase fluctuations of the Raman lasers and observing that these are flat and unstructured and have no features that would lead to memory effects on the few microsecond timescale. Turning to heating, the heating rate observed in QSCOUT is on the order of a phonon per millisecond (this depends on the motional mode of concern, but is an accurate order of magnitude value) \cite{PhysRevA.91.041402}, and hence, over the course of such long-time driven dynamics over $5000 \mu$s, it is likely that heating effects become observable. To investigate the impact of heating and motional dynamics on the learned NMZ models we collected additional datasets at different axial trap frequencies. The trap frequency dictates frequency of ion motion and moreover, the rate of heating of ions is dependent on the motional frequency. As before, there were ten time traces from random initial states for each axial trap frequency, and we learned NMZ models using the same approach as above. In each case, we fit an exponential decay to the norm of the NMZ operators as a function of memory kernel length, \ie $||\hat{\bm{\Omega}}_{\Delta}^{(l)}||_2 = A e^{-\gamma_m l\Delta} + B$, and use the fit decay parameter, $\gamma_m$, as a measure of the length of memory kernel required to capture the data.
Table \ref{tab:qscout_axial_table} shows the fit decay parameter for each of the axial trap frequencies, and there is a clear trend of slower decay (longer memory kernel) for increasing trap frequency. This is indirect evidence that motional dynamics and heating are correlated with the degree of non-Markovianity of the long-time driven dynamics. While this analysis provides an interesting lead, further experiments and analysis are required to clearly isolate the source(s) of noise responsible for the observed non-Markovianity.

 \begin{table}[h!]
 \centering
 \begin{tabular}{ | c | c | }
 	\hline
 	Axial Trap Frequency (MHz) & Exponential Decay Constant $\gamma_m$ ($\mu \text{s}^{-1}$) \\
	\hline
	0.35 & -1.01\\
	\hline
	0.45 & -0.95\\
	\hline
	0.55 & -0.68 \\
	\hline
 \end{tabular}
 \caption{Table showing how axial trap frequency affects the memory kernel length of the best-fit NMZ models in the long-time driven dynamics experiments.  A more negative decay constant indicates the memory kernel decays to zero more quickly, so there is a clear trend that as the axial trap frequency increases, the memory kernel length also increases.}
 \label{tab:qscout_axial_table}
 \end{table}

\subsubsection*{Pauli-X Gate}
        In a quantum computing setting, ions will not be continuously driven for timescales as long as 5000 $\mu$s. Instead, short duration quantum gates are applied during the execution of a quantum algorithm. Therefore, we now study the dynamics of a trapped-ion qubit during the execution of a Pauli-X gate, which is ideally a rotation about the $\sigma_x$ axis by an angle of $\pi$ radians on the Bloch sphere. This corresponds to driving of the ion with Raman pulses for $T=25$ $\mu$s. Therefore, we now learn the dynamics with time series of this length. We prepare the qubit in 30 random initial states, turn on the drive for $T=25$ $\mu$s, and measure the Bloch vector every 1.0 $\mu$s using 400 shots per Pauli expectation. We train the NMZ model using LOOCV with these 30 time series.

		\begin{figure*}
			\begin{subfigure}{0.3\textwidth}
				\includegraphics[width=\textwidth]{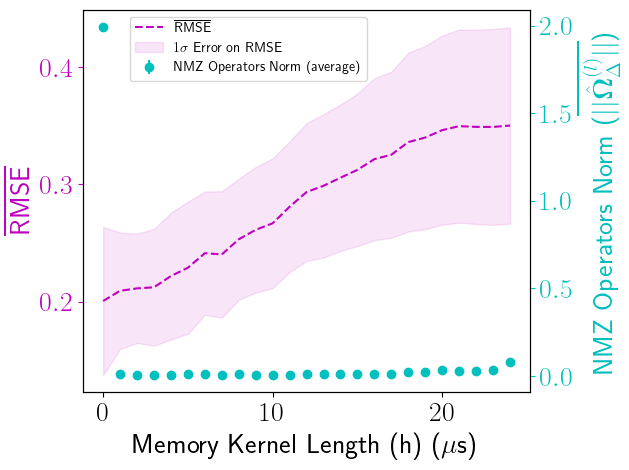}
				\label{fig:qscout_markov_error_plot}
				\caption{}
			\end{subfigure}
			\begin{subfigure}{0.7\textwidth}
				\hspace{10pt}
				\includegraphics[width=\textwidth]{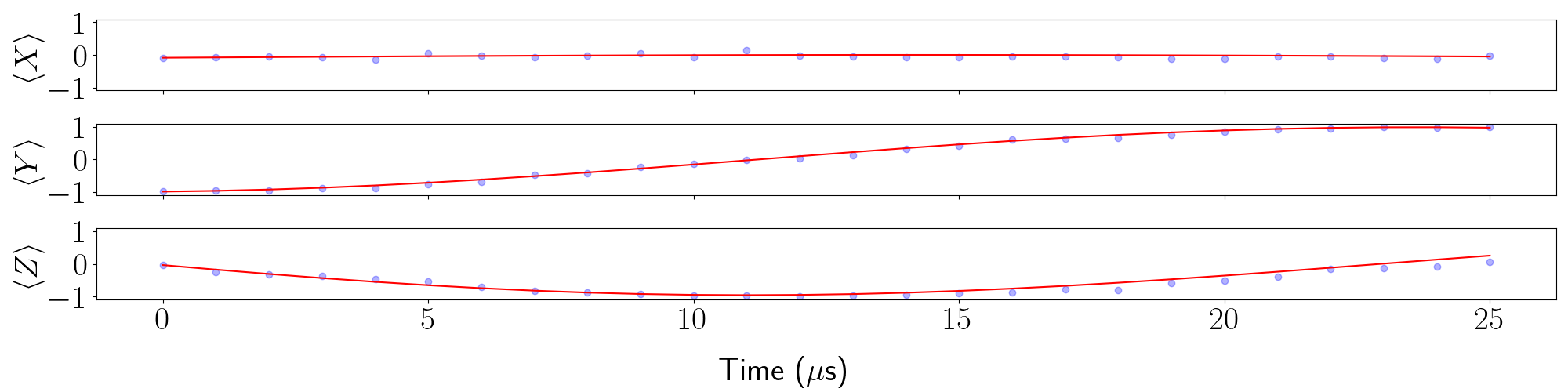}
				\vspace{5pt}
				\label{fig:qscout_markov_comparison_plot}
				\caption{}
			\end{subfigure}
			\caption{\label{fig:qscout_markov} A single qubit undergoes a Pauli-X rotation of angle $\pi$ on the Bloch sphere, or equivalently a Pauli-X gate. \textbf{(a)} $\overline{\text{RMSE}}$ (dashed magenta line) and average NMZ operator norms (cyan markers) for the learned NMZ models as a function of memory kernel length. \textbf{(b)} Example of predictions of the Bloch vector evolution during the gate execution for one of the learned models given the initial state for one of the time series. The blue points are the measured Pauli expectations, and the red lines are the predictions.}		\end{figure*}

        \cref{fig:qscout_markov} shows the predictive performance of the learned model. \cref{fig:qscout_markov}(a) indicates the model performance (in terms of RMS predictive error) remains constant to within error bars for small memory kernels and then degrades quickly with longer memory kernel lengths. The parsimonious interpretation of this is that the dynamics is Markovian at the gate timescale. \cref{fig:qscout_markov}(b) shows the excellent prediction accuracy of a Markov model (one of the learned NMZ models, restricted to just the Markov matrix, $\bm{\Omega}_{1.0 \mu s}^{(0)}$).
        The learned discrete-time Markov transition matrix from one of the NMZ models takes the form: 
		\begin{align}
			\small
			\hat{\bm{\Omega}}_{1.0 \mu s}^{(0)} =
			\begin{pmatrix}
				0.9957 & -0.0027 & 0.032 & -0.021 \\
				9.0 \times 10^{-4} & 0.9936 & -8.3 \times 10^{-3} & -1.7 \times 10^{-3} \\
				9.6 \times 10^{-3} & -1.0 \times 10^{-3} & 0.9911 & -0.124 \\
				-7.4 \times 10^{-3} &	-7.1 \times 10^{-3} &	0.135 & 0.9810
			\end{pmatrix}
			\label{eqn:qscout_markov_omega0}
		\end{align}
        We can interpret this learned Markov matrix by comparing to the parameterized form of a general discrete-time Markov transition matrix, \cref{eqn:markov_generator}. Assuming $\gamma_+=0$ (which is reasonable since incoherent excitation is very unlikely in this system), we can estimate the values of the other physical rates from this matrix by algebraically solving a set of equations. This yields: $\omega_x \approx 0.0675/\Delta, \omega_y \approx 0, \omega_z \approx 0$, indicating that the Hamiltonian is primarily Pauli-X. $\gamma_- \approx 0.0074/\Delta$, meaning incoherent relaxation (T1 process) is not negligible on these timescales. $\Gamma_z \approx 0, \Gamma_y \approx 0.002275/\Delta$, and $\Gamma_x \approx 0.003525/\Delta$, implying that there is dephasing along the $\sigma_x$ and $\sigma_y$ axes, which is likely due to broadband intensity and phase fluctuations of the driving lasers. These interpretations of the learned operators illustrate the utility of our approach to diagnosing quantum gates.

		\section*{Discussion}
		We have demonstrated that data-driven learning of open system dynamics of qubits using the NMZ representation is tractable and useful. The learned models are interpretable and particularly useful for inferring the non-Markovianity of the dynamics, something that has been difficult to do in generality using other existing methods. The regression-based learning algorithm we use is stable and efficient in the sense that it requires relatively little computational time and training data compared to deep learning models for the same task. By demonstrating our approach on experimental data, we have shown its applicability to real-world noisy data settings. 
        
        While the demonstrations in this work have been for single-qubit systems, the extension to the characterization of multi-qubit gates is straightforward. The computational expense and amount of training data is likely to increase for multi-qubit systems -- to completely characterize $n$-qubit dynamics requires tracking $4^n-1$ observables. However, all practical quantum computing platforms utilize native gates that operate on one or two qubits and our approach will remain tractable in these settings. 

        The focus of this work has been to provide the first application of NMZ-based learning to qubit dynamics. Given the successes demonstrated here, there are several natural avenues for future work. Firstly, it would be interesting to study whether our learning approach can be made \emph{self-consistent} \cite{PhysRevA.87.062119}. Loosely, this property requires removing assumptions of ideality from any device used to collect characterization data. In this work we have assumed that the state preparation and measurement processes are ideal, and this should be relaxed for self-consistent characterization. Secondly, while the regression-based algorithm for data-driven learning of the NMZ equation is optimal in terms of minimizing a mean-squared error in the noiseless setting \cite{lin2023}, it would be interesting to study other learning approaches, such as those based on neural networks, to determine if they present any advantages in the noisy data setting. The noise-sensitivity of the regression-based learning method is evidenced in the $\overline{\text{RMSE}}$ peaks in \cref{fig:qscout}(a). While in this case the $\overline{\text{RMSE}}$ peaks occurred at memory kernel lengths that are clearly irrelevant (\ie at values of $h > h^\star$), it does motivate other learning methods that are more noise-robust. Lin \emph{et al.} have already demonstrated a learning method that treats the NMZ operators as neural networks and showed that this method can improve the predictive power of the NMZ model \cite{lin2023}.  However, this comes at the cost of turning the NMZ operators into black boxes with little interpretability.  A workaround for this could be to use the linear regression technique to only learn $\bm{\Omega}_{\Delta}^{(0)}$ and then switch to the neural network models to learn the remaining operators.  This way, the interpretability of the Markovian dynamics are preserved, yet some of the improved predictive power of the neural networks is gained.  It could also be possible to just construct machine learning models for all the operators that have the improved predictive power of neural networks while maintaining interpretability, similar to the STEADY method \cite{steady}.

		\section*{Methods}
		\subsection*{Learning the Discretized NMZ Equation}
		The NMZ equation is an integro-differential equation that describes any open system exactly with no approximations \cite{nmz}.
		
		\begin{equation}
			\frac{d}{dt}\mathbf{g}(t) = \mathbf{M}\cdot\mathbf{g}(t) - \int_0^t \mathbf{K}(s)\cdot\mathbf{g}(t-s)ds + \mathbf{F}(t),
            \label{eq:NMZ_cont}
		\end{equation}
		where $\mathbf{g}(t)$ is a vector of all chosen observables of the system at time \textit{t}, $\mathbf{M}$ is the \textit{Markov transition matrix}, which describes the Markovian dynamics of the system.  $\mathbf{K}(t)$ is the \textit{memory kernel}, which encodes how previous states of the system will affect evolution of the current state, and $\mathbf{F}(t)$ is the generalized Langevin noise term, which fundamentally stems from uncertainty about the initial state of the degrees of freedom not capture in $\mathbf{g}(t)$. There are many textbook derivations of the NMZ equation in the context of open quantum system dynamics, and we refer the reader to the treatment in Breuer and Petruccione \cite[Chapter 9]{bp} for details. 

        The degrees of freedom tracked in $\mathbf{g}(t)$ are arbitrary, but in our setting, the natural choice is a complete set of observables for the qubits being characterized. In general, non-unital quantum dynamics can have autonomous forcing terms and hence, the vector $\mathbf{g}(t)$ should be supplemented with a constant term as done in the main text. 

        Data-driven learning of the NMZ equation proceeds by collecting training data in the form of $N$ time series $\mathbf{g}(t)$, $t_0 \leq t \leq T$, for $M$ different initial conditions. This data is always sampled from the real time evolution and therefore discrete. While it is possible to learn the operators in the original NMZ equation, \cref{eq:NMZ_cont}, from this data using finite difference approximations, a more numerically stable approach is to formulate a time-discretized version of the NMZ equation \cite{lin2021, lin2023}
		
		\begin{equation}
			\mathbf{g}((k+1)\Delta) = \sum_{l=0}^k \bm{\Omega}_{\Delta}^{(l)} \cdot \mathbf{g}((k-l)\Delta) + \mathbf{W}_k,
		\end{equation}
		where $\Delta$ is the time interval between sampled points in the trajectory, $k$ is a timestep index such that $t = k\Delta$, $\bm{\Omega}_{\Delta}^{(l)}$ is the $l$-th order NMZ operator that encodes $\mathbf{M}$ or $\mathbf{K}$ in the discretized form of the NMZ equation, and $\mathbf{W}_k$ is the discretized form of $\mathbf{F}(t)$.  For $\Delta \ll 1$, we can approximate 
        \begin{align}
            \bm{\Omega}_{\Delta}^{(0)} &\approx \mathbf{I} + \mathbf{M}\Delta \label{eq:M_approx}\\
            \bm{\Omega}_{\Delta}^{(l)} &\approx \Delta^2\mathbf{K}(l\Delta), \quad \text{for} \quad l>1 \label{eq:K_approx}
        \end{align}

        Lin \emph{et al.} have formulated a regression-based learning algorithm for the operators that enter the discretized NMZ equation \cite{lin2021,lin2023}, which is what we use in the main text to demonstrate learning of qubit dynamics. We briefly summarize this algorithm here. 
		We assume all $N$ time series that form the training data are sampled with the same $\Delta$ and for the same duration, $T$. This data can be organized into an $N \times M \times K$ matrix called $\mathbf{X}$, where $M$ is the number of chosen observables and $K$ is the number of timesteps in the sampled trajectories.  Both $M$ and $K$ must be the same across all $N$ trajectories.  We denote $\mathbf{X}[k]$ as the $N\times M$ matrix that contains the values of all $M$ observables across all $N$ time series, at time $k\Delta$.
        
        Once $\mathbf{X}$ is created, the empirical $k$-lag correlation matrix of the observables needs to be calculated, 
		\begin{equation}
			\mathbf{C}[k] = \frac{1}{N(h^\star-k)}\sum_{l=0}^{h^\star-k-1}\mathbf{X}^T[k+l]\cdot \mathbf{X}[l], \quad k\in [0,h^\star-1],
		\end{equation}
        where $h^\star$ is a memory kernel length cut-off parameter, which can be as large as $K$.
        From here, it is possible to compute the discrete-time Markov transition matrix, $\bm{\Omega}^{(0)}_{\Delta}$:
		\begin{equation}
			\bm{\Omega}^{(0)}_{\Delta} = \mathbf{C}[1]\cdot \left(\mathbf{C}[0]\right)^{-1}
		\end{equation}
		
		Once $\mathbf{C}(k\Delta)$ and $\bm{\Omega}^{(0)}_{\Delta}$ are known, the remaining $\bm{\Omega}^{(l)}_{\Delta}$ can be computed iteratively,
		\begin{equation}
			\bm{\Omega}^{(n+1)}_{\Delta} = [\mathbf{C}[n+2] - \sum_{l=0}^n \bm{\Omega}^{(l)}_{\Delta}\cdot\mathbf{C}[n-l+1]]\cdot \left(\mathbf{C}[0]\right)^{-1}
		\end{equation}
		
		The only discretized operators left to learn are $\mathbf{W}_k$.  While there is a method to learn the orthogonal dynamics, it is limited to the initial state of the time evolution it is trained from \cite{lin2021,lin2023}.  In other words, the $\mathbf{W}_k$ depends on which time evolution it was learned from, so it cannot be applied to time evolutions outside the training dataset.  Previous work has attempted to model the orthogonal dynamics \cite{lin2023} or assume $\mathbf{W}_k = 0$ \cite{lin2021, lin2023}.  In the open quantum systems setting, as shown in \cite[Chapter 9]{bp}, if one assumes that the initial state of the system of interest (the qubits in our case) and its environment is factorizable, $\rho_{total}(0) = \rho_{qubits}(0) \otimes \rho_{env}(0)$, for some fixed but arbitrary environment state $\rho_{env}(0)$, then the generalized Langevin noise term is zero. This factorizable initial state assumption is typically a good one in the quantum computing setting, where characterization experiments begin with qubits being reset to a fiducial state that is pure, and hence, we assume  $\mathbf{W}_k = 0$ always. 

        Once the operators in the discrete-time NMZ equation have been learned, assuming $\Delta \ll 1$, we can estimate the equivalent operators in the continuous-time NMZ equation (using \cref{eq:M_approx,eq:K_approx}), which are often more amenable to physical interpretation. As discussed by Lin \emph{et al.} in the noiseless setting, this regression-based framework for data-driven learning of the NMZ equation is optimal in terms of minimizing a mean-squared error \cite{lin2023}.

        \subsection*{Trapped-Ion Hardware Data Generation}
        In the QSCOUT system \cite{qscout_2021} linear chains of $^{171
}$Yb$^+$ qubits are confined above a microfabricated surface trap with radio frequency and direct current control voltages. A 370 nm laser is used for Doppler cooling, detecting , and preparation of the qubit in a subspace spanned by the hyperfine states $^2S_{1/2}\ket{F=0, m_F=0}$ ($\ket{0}$) and $^2S_{1/2}\ket{F=1, m_F=0}$ ($\ket{1}$), split by 12.6 GHz.  The frequency comb of a pulsed 355 nm laser is used to drive Raman transitions on the qubit carrier transition (using copropagating beams) as well as motional sidebands for sideband cooling. See reference \citeonline{qscout_2021} for more details of the QSCOUT system.

\section*{Acknowledgments}
We thank Matthew N. H. Chow for providing early versions of the experimental data, and Brandon Ruzic for discussions about noise sources in trapped ion experiments.
This work was supported by the US Department of Energy, Office of Science, Office of Advanced Scientific Computing Research through its Accelerated Research in Quantum Computing Program MACH-Q Project and its Quantum Testbed Program. Sandia National Laboratories is a multimission laboratory managed and operated by National Technology and Engineering Solutions of Sandia LLC, a wholly owned subsidiary of Honeywell International Inc. for the U.S. Department of Energy’s National Nuclear Security Administration under contract DE-NA0003525.
		
		\bibliography{main}
			
\end{document}


\maketitle

	\section{Simulation: non-Markovian evolution}
	
	In this section, we look at the matrix elements of the NMZ operators for the $\gamma=0.5$ (Fig. 1a) and $\gamma=10$ (Fig. 1b) data.  In both cases, all matrix elements quickly decay to zero regardless of their values at zero memory kernel length, which indicates the memory kernels can be truncated without losing model fidelity.

	\begin{figure*}
		\begin{subfigure}{\textwidth}
			\includegraphics[width=\textwidth]{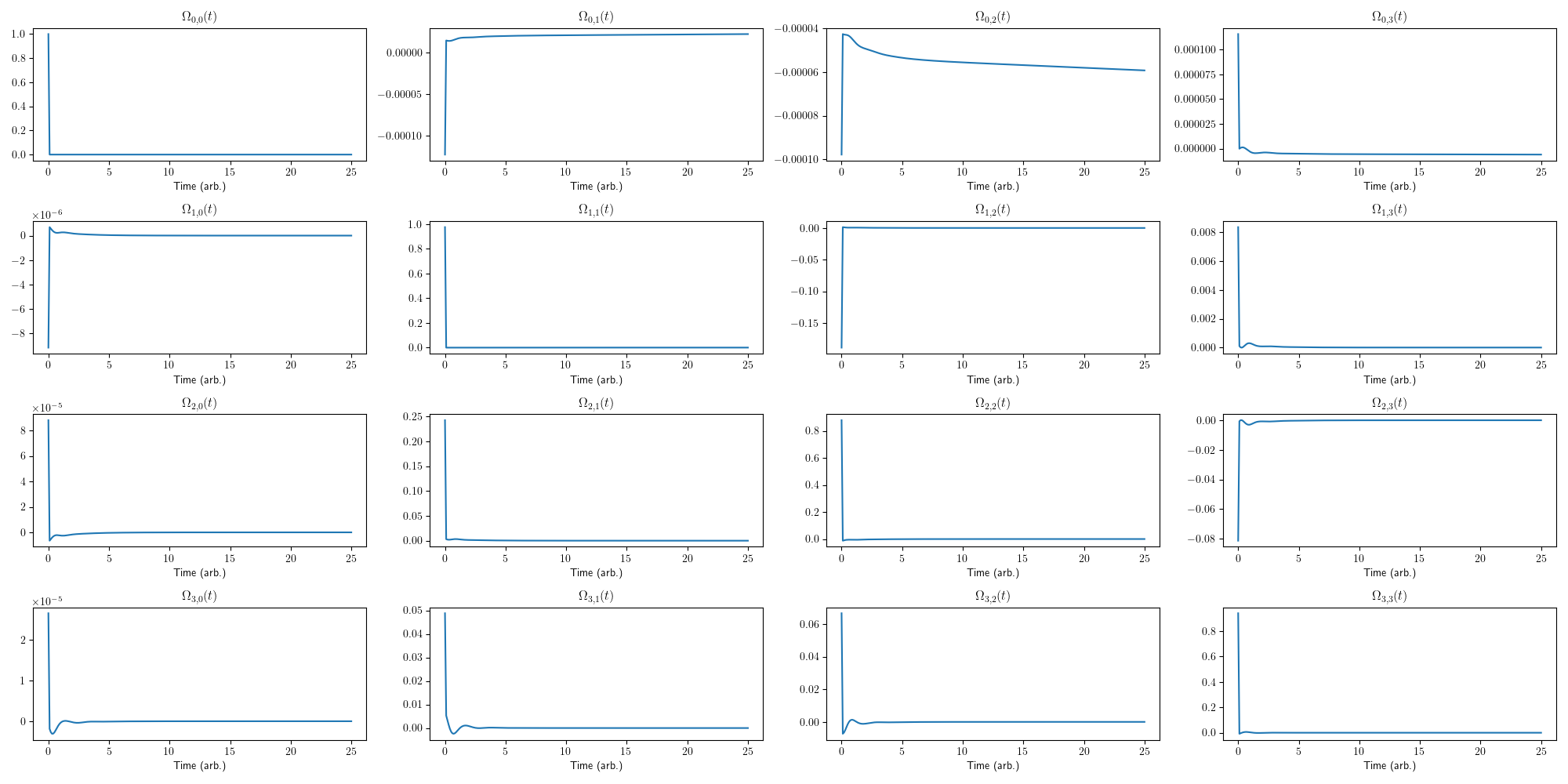}
			\label{fig:omega_elements_gamma_05}
			\caption{}
		\end{subfigure}
		\begin{subfigure}{\textwidth}
			\includegraphics[width=\textwidth]{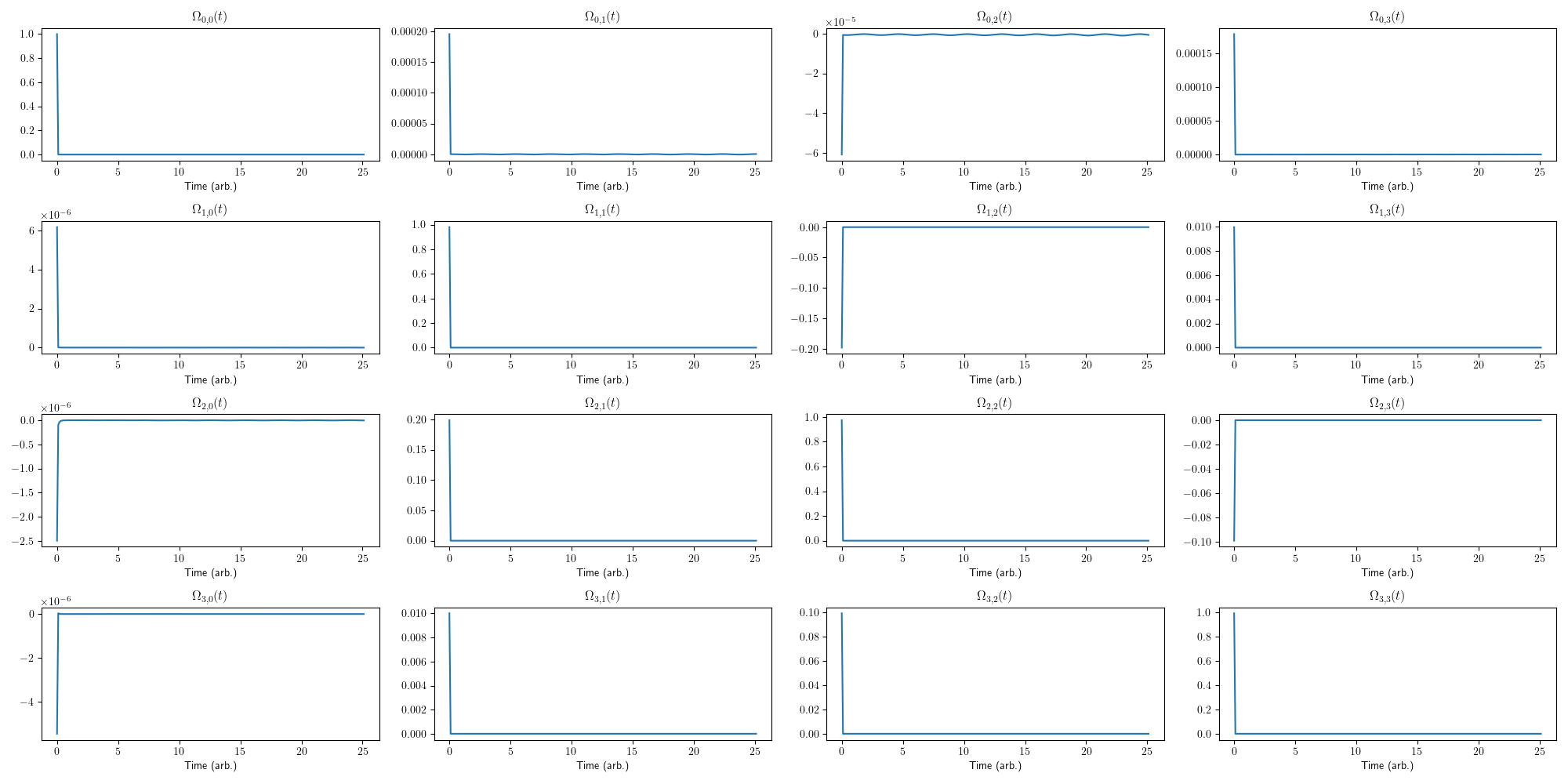}
			\label{fig:omega_elements_gamma_10}
			\caption{}
		\end{subfigure}
		\label{fig:ou_extended}
		\caption{Additional figures for the non-Markovian simulation section. \textbf{(a)} shows matrix elements of the learned NMZ operators for the $\gamma = 0.5$ dynamics as a function of memory kernel length. \textbf{(b)} shows matrix elements of the learned NMZ operators for the $\gamma = 10.0$ dynamics as a function of memory kernel length.}
	\end{figure*}

	\section{Experiment: Long-time driven dynamics}
	
	We cover additional figures in this supplementary section to give added details to the long-time driven dynamics in the trapped-ion hardware experiments. Fig. 2(a) shows how, unlike the simulated data, the matrix elements oscillate about zero after decaying with increasing memory kernel length.  These oscillations are caused by shot noise in the measured data and it is suspected these oscillations would be smaller in amplitude with a larger number of shots.  These oscillations are still quite small however, indicating the memory kernel length can be truncated without losing model predictive accuracy.  Fig. 2(b) compares the performance of a learned Markovian model (top) to a learned NMZ model with the optimized memory kernel length truncation of $h^\star = 22.5$ $\mu$s (bottom) on test data unseen in the training phase.  The Markovian model does not capture the Rabi oscillations and seems to be consistent with a Markovian model with large jump operators causing the qubit to decay to its steady state very quickly.  The non-Markovian model is able to capture the Rabi oscillations in the qubit and the rate at which the qubit decays to its steady state.   This huge improvement in model performance indicates the driven qubit is a non-Markovian system.
	
	\begin{figure*}
		\begin{subfigure}{\textwidth}
			\includegraphics[width=\textwidth]{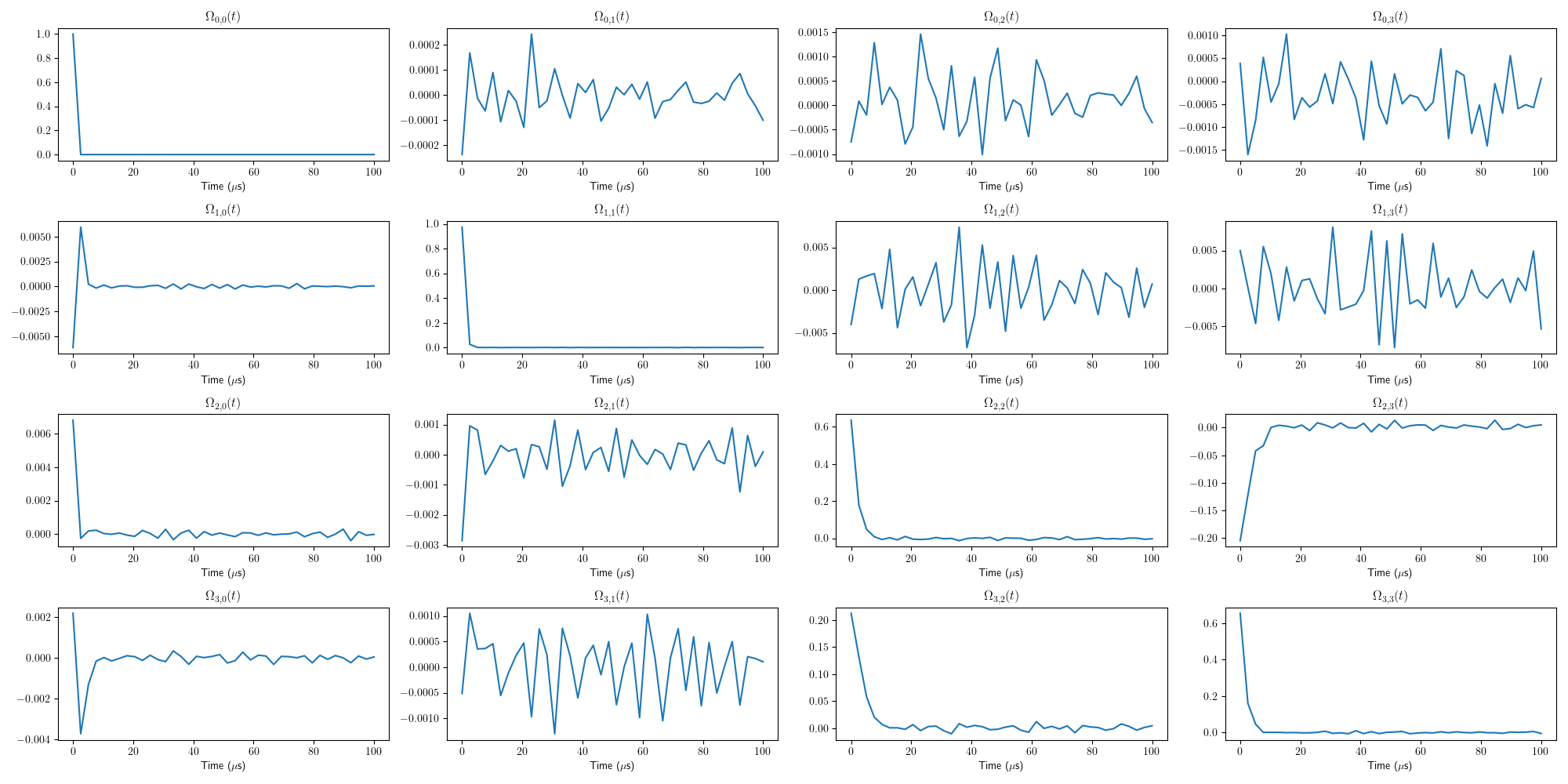}
			\caption{}
		\end{subfigure}
		\begin{subfigure}{\textwidth}
			\includegraphics[width=\textwidth]{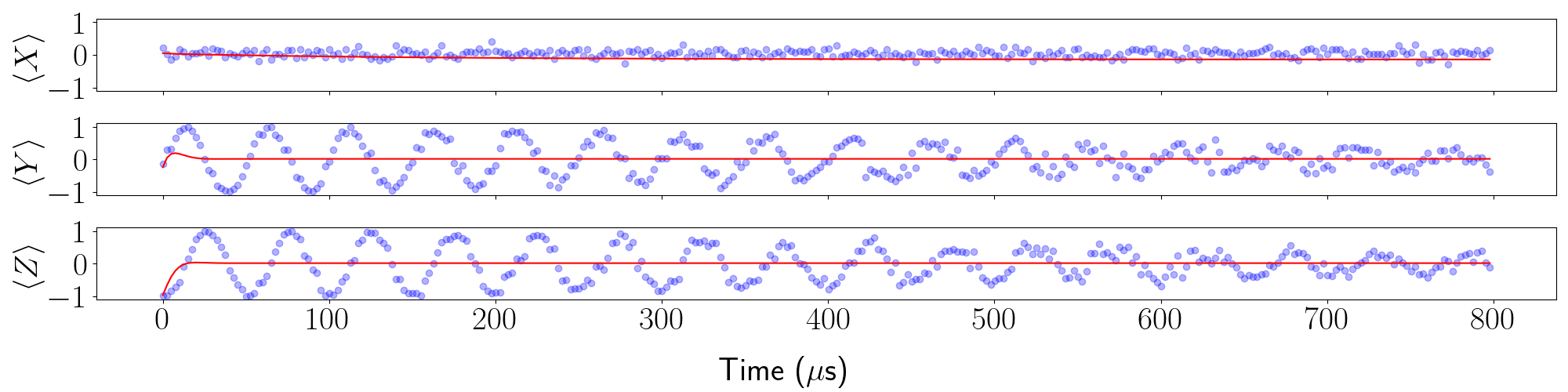}
			\includegraphics[width=\textwidth]{qscout/plot_comparison_test_state_0_mu_0_h_9.png}
			\caption{}
		\end{subfigure}
		\label{fig:qscout_extended}
		\caption{Additional figures for QSCOUT long-range dynamics section. \textbf{(a)} Matrix elements of the learned NMZ operators as a function of memory kernel length. \textbf{(b)} Comparison of predictions of the learned NMZ model for an unseen initial state with different memory kernel lengths. The red lines in the top set of plots are the predictions of the learned NMZ model with zero memory kernel length (Markovian).  The red lines in the bottom set of plots are the predictions of the learned NMZ model with memory kernel length truncation of $h^\star = 22.5$ $\mu$s.}
	\end{figure*}
	
	Fig. 3 focuses on comparing the three experiments with different axial trap frequencies.  \textbf{(a)}-\textbf{(c)} show the $\overline{\text{RMSE}}$ and NMZ operator norms as a function of memory kernel length for the 0.35, 0.45, and 0.55 MHz trap frequency experiments respectively.  These plots also contain exponential decay fits to the errors and operator norms that allow us to more easily see the behavior of the error and operators norms as a function of memory kernel length.  \textbf{(d)} is a table containing the exponential decay fits of the error and operators norms for each axial trap frequency.
	
	\begin{figure*}
		\begin{subfigure}{0.3\textwidth}
			\includegraphics[width=\textwidth]{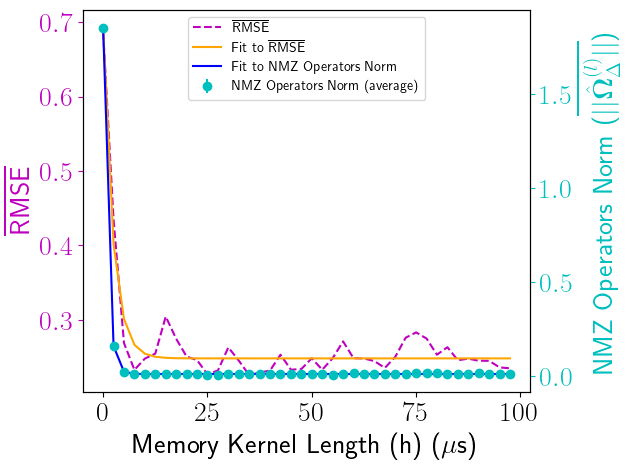}
			\caption{}
		\end{subfigure}
		\begin{subfigure}{0.3\textwidth}
			\includegraphics[width=\textwidth]{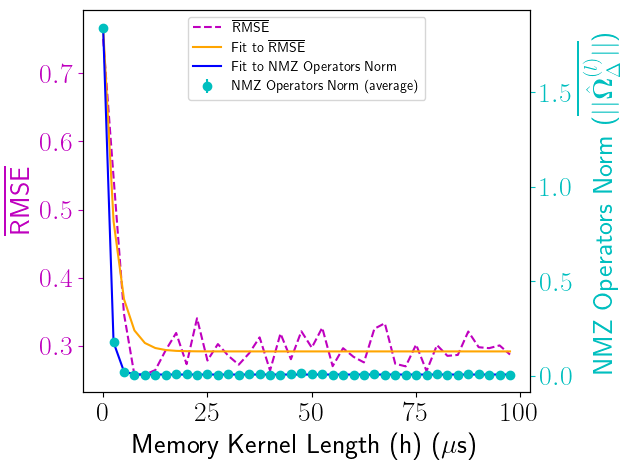}
			\caption{}
		\end{subfigure}
		\begin{subfigure}{0.3\textwidth}
			\includegraphics[width=\textwidth]{qscout/fit_to_error_plot.png}
			\caption{}
		\end{subfigure}
		\begin{subfigure}{\textwidth}
			\begin{center}
			\begin{tabular}{ | c | c | c | }
				\hline
				Axial Trap Frequency (MHz) & $\overline{\text{RMSE}}$ Fit & NMZ Operators Norm Fit \\
				\hline
				0.35 & $\overline{\text{RMSE}} = 0.44e^{-0.43 (2.5 \mu s)l}+0.25$ & $|| \hat{\bm{\Omega}}_{2.5 \mu s}^{(l)} || = 1.84e^{-1.01 (2.5 \mu s)l}+0.01$ \\
				\hline
				0.45 & $\overline{\text{RMSE}} = 0.48e^{-0.36 (2.5 \mu s)l}+0.29$ & $|| \hat{\bm{\Omega}}_{2.5 \mu s}^{(l)} || = 1.83e^{-0.95 (2.5 \mu s)l}+0.007$ \\
				\hline
				0.55 & $\overline{\text{RMSE}} = 0.42e^{-0.28 (2.5 \mu s)l}+0.38$ & $|| \hat{\bm{\Omega}}_{2.5 \mu s}^{(l)} || = 1.67e^{-0.68 (2.5 \mu s) l}+0.01$ \\
				\hline
			\end{tabular}
			\end{center}
			\caption{}
		\end{subfigure}
		\label{fig:qscout_more_trap_frequencies}
		\caption{The $\overline{\text{RMSE}}$ of the learned NMZ model predictions and the L2 norms of the learned NMZ operators as a function of memory kernel length for three different axial trap frequencies.  \textbf{(a)}-\textbf{(c)} The $\overline{\text{RMSE}}$ and NMZ operator norms for an axial trap frequency of 0.35, 0.45, and 0.55 MHz respectively. \textbf{(d)} A table showing the exponential decay fits for the $\overline{\text{RMSE}}$ and NMZ Operators Norms as a function of memory kernel length in $\mu$s for each experiment with different axial trap frequencies.}
	\end{figure*}